\documentclass{article}
\usepackage{epsfig}
\usepackage{amsthm}
\usepackage{amsmath}
\usepackage{amssymb}
\usepackage{amsbsy}
\usepackage{amscd}
\usepackage{color}
\usepackage{graphicx}
\usepackage[colorlinks=true, pdfstartview=FitV, linkcolor=blue, citecolor=blue, urlcolor=blue]{hyperref}

\usepackage{verbatim}
\DeclareMathOperator{\sgn}{sgn}

\makeatletter
\@addtoreset{equation}{section}
\makeatother

\def\bc{\begin{corollary}}
\def\ec{\end{corollary}}
\def\&{&{\hskip -20pt}}

\def\br{\begin{remark}\rm\small}
\def\1{{\bf 1}}
\def\er{\end{remark}}
\def\bt{\begin{theorem}}
\def\et{\end{theorem}}

\def\bx{\begin{examp}}
\def\ex{\end{examp}}
\def\bd{\begin{definition}}
\def\ed{\end{definition}}
\def\bp{\begin{proposition}\rm}
\def\bl{\begin{lemma}\em}
\def\el{\end{lemma}}
\def\ep{\end{proposition}}
\def\bea{\begin{eqnarray}}
\def\eea{\end{eqnarray}}

\def\C{{\mathbb C}}
\def\R{{\mathbb R}}

\newtheorem{theorem}{Theorem}[section]
\newtheorem{examp}{Example}[section]
\newtheorem{coroll}{Corollary}[section]
\newtheorem{examps}{Examples}[section]

\newtheorem{lemma}{Lemma}[section]
\newtheorem{remark}{Remark}[section]
\newtheorem{remarks}[remark]{Remarks}
\newtheorem{proposition}{Proposition}[section] 
\newtheorem{definition}{Definition}[section]
\def\br{\begin{remark}}
\def\er{\end{remark}}
\def\bt{\begin{theorem}}
\def\et{\end{theorem}}
\def\bc{\begin{coroll}}
\def\ec{\end{coroll}}
\def\brs{\begin{remarks} \rm\
\begin{enumerate}}
\def\ers{\end{enumerate}\end{remarks}}
\def\bl{\begin{lemma}}
\def\el{\end{lemma}}
\def\bxs{\begin{examps}. \rm\begin{enumerate}}
\def\exs{\end{enumerate}\end{examps}}
\def\bd{\begin{definition}}
\def\ed{\end{definition}}
\def\bp{\begin{proposition}}
\def\ep{\end{proposition}}
\def\be{\begin{equation}}
\def\ee{\end{equation}}
\def\bew{\begin{equation*}}
\def\eew{\end{equation*}}

\def\d{{\rm d}}
\def\bea{\begin{eqnarray}}
\def\eea{\end{eqnarray}}
\def\beas{\begin{eqnarray*}}
\def\eeas{\end{eqnarray*}}

\def\iint{\int\!\!\!\!\int}
\def\C{{\mathbb C}}

\def\R{{\mathbb R}}

\date{}
\begin{document}
%

\baselineskip 16pt plus 1pt minus 1pt
\begin{titlepage}
\begin{flushright}
\end{flushright}
\vspace{0.2cm}
\begin{center}
\begin{Large}
\textbf{ Cubic String Boundary Value Problems and Cauchy Biorthogonal Polynomials}
\end{Large}\\
\bigskip
\begin{large} {M.
Bertola $^{\dagger\ddagger}$ \footnote{Work supported in part by the Natural
    Sciences and Engineering Research Council of Canada (NSERC),
    Grant. No. 261229-03 and by the Fonds FCAR du
    Qu\'ebec No. 88353.},  M. Gekhtman~$^a$ 
\footnote{Work supported in part by NSF Grant DMD-0400484.}, J. Szmigielski~$^b$ \footnote{Work supported in part by the Natural
    Sciences and Engineering Research Council of Canada (NSERC),
    Grant. No. 138591-04}}
\end{large}
\\
\bigskip
\begin{small}
$^{\dagger}$ {\em Centre de recherches math\'ematiques,
Universit\'e de Montr\'eal\\ C.~P.~6128, succ. centre ville, Montr\'eal,
Qu\'ebec, Canada H3C 3J7 \\
~~E-mail: bertola@crm.umontreal.ca}
\smallskip

$^{\ddagger}$ {\em Department of Mathematics and
Statistics, Concordia University\\ 7141 Sherbrooke W., Montr\'eal, Qu\'ebec,
Canada H4B 1R6} \\
\smallskip
~$^a$ {\em Department of Mathematics
255 Hurley Hall, Notre Dame, IN 46556-4618, USA\\
~~E-mail: Michael.Gekhtman.1@nd.edu}

\smallskip
~$^b$ {\em Department of Mathematics and Statistics, University of Saskatchewan\\ 106 Wiggins Road, Saskatoon, Saskatchewan, S7N 5E6, Canada\\
~~E-mail: szmigiel@math.usask.ca}
\end{small}
\end{center}
\bigskip
\begin{abstract}
Cauchy Biorthogonal Polynomials appear in the study of special solutions to the 
dispersive nonlinear partial differential equation called the Degasperis-Procesi (DP) equation,  
as well as in certain two-matrix random matrix models.  Another context in which 
such biorthogonal polynomials play a role is the cubic string; 
a third order ODE boundary value problem $-f'''=zg f$ which is a generalization of the inhomogeneous string problem studied by M.G. Krein.  
A general class of such boundary value problems 
going beyond the original cubic string problem associated with the DP equation is discussed
under the assumption that the source of inhomogeneity $g$ is a discrete measure.  
It is shown that by a suitable choice of a generalized Fourier transform associated to 
these boundary value problems one can establish a Parseval type identity which 
aligns Cauchy biorthogonal polynomials with certain natural orthogonal systems on $L^2_g$.  
\end{abstract}
\medskip
\bigskip
\bigskip
\bigskip
\bigskip

\end{titlepage}

\section{An ordinary inhomogeneous string, the Degasperis-Procesi equation and a cubic string}\label{sec:cubic string}
 \subsection{DP equation}
 The vibrating string is one of the most fundamental physical phenomena, whose 
 mathematical foundations go back at least to the times of D. Bernoulli 
 with important contributions of J.B. Fourier, B. Riemann, S. Sobolev and M.G. Krein and it appears that the 
 heated discussion of more refined aspects of the problem has not subsided.  Of multiple reasons 
 for the special role of the vibrating string two stand out: 1) this is the simplest problem with infinitely many degrees of freedom and as such it inevitably leads to questions of analysis on function 
 spaces, in particular the eigenvalue problems for infinitely large matrices; 2) the vibrating string problem is a prototype of hyperbolic problems 
 and plays an important role in basic mathematical education.  
 In its simplest, one dimensional, form, one studies the partial differential equation 
 $$ 
 \frac{1}{c^2} u_{tt}-u_{xx}=0,   \qquad 0<x< 1, 
 $$ 
 for the amplitude of the string $u(x,t)$.  The coefficient $\frac{1}{c^2}$ is proportional to the mass density $\rho $ of the string and in principle it does not have to be homogeneous (constant) in $x$.  Thus, in general, 
 $\frac{1}{c^2}=\frac{\rho(x)}{T}$, where $T$ represents the tension in the string and 
 $\rho(x)$ is the linear mass density of the string. Then the fundamental modes, $u(x,t)=v(x)\cos(\omega t)$, are given by solving: 
 \begin{equation*}
 -v_{xx}=\frac{\omega^2}{T} \rho(x) v,   \qquad 0<x<1, 
 \end{equation*}
 subject to some boundary conditions, corresponding physically to how the string is 
 tied.  The most common are the Dirichlet boundary conditions, $v(0)=v(1)=0$ (the string is tied on both ends), and 
 the Neumann conditions, $v_x(0)=v_x(1)=0$ (the end points can slide in the vertical direction only).  Once equipped with the boundary conditions, say the Dirichlet conditions, the problem is turned into an 
 eigenvalue problem of the Sturm-Liouville type:
 \begin{equation}\label{eq: Kstring}
 -v_{xx}=z \rho(x) v,   \qquad v(0)=v(1)=0, \qquad 0<x<1
 \end{equation}
 where $z=\frac{\omega^2}{T}$.  
 Such an inhomogeneous string problem was studied in an influential work by M.G. Krein
 \cite{Kreinstring} in the 1950s as a generalization of Stieltjes' theory of analytic 
 continued fractions \cite{KreinStieltjes}.  
 
 The present paper addresses several aspects of Krein's theory for 
 a third order equation: 
 $$ 
 -\phi_{xxx}=z \rho \phi,  \qquad 0<x<1
 $$
 subject to some boundary conditions specified later.  We will refer to this equation as 
 the \emph{ cubic string}.    
 Since any third order problem is non-self adjoint one can seriously doubt any applicability of 
 this type of equation to physical systems.  Below we briefly sketch how the cubic string has 
 arisen in a study of certain nonlinear partial differential equation 
 modeling weekly dispersive waves.  The equation in question is the Degasperis-Procesi (DP) equation \cite{dp}:
 \begin{equation}
  \label{eq:DPsingle}
  u_t - u_{xxt} + 4 u u_x = 3 u_x u_{xx} + u u_{xxx}, \qquad (x,t)\in \mathbb{R}^2,
\end{equation}
where $u(x,t)$ is the wave hight at $x$ and time $t$.  The DP equation 
admits a Lax formulation, first proposed in \cite{dhh1}, which means that 
the DP equation follows from the compatibility condition for the 
the system 
\begin{subequations}
  \label{eq:lax}
  \begin{align}
    \label{eq:lax1}
    (\partial_x - \partial_x^3) \psi &= z \, m\psi, \\
    \label{eq:lax2}
   \partial_t  \psi &= \left[ z^{-1} (1-\partial_x^2) + u_x - u \partial_x \right] \psi.
  \end{align}
\end{subequations}
where $z\in \C$.  In general $m$ can be a Radon measure, for example 
$m$ can be a discrete measure $m=\sum_{i=1}^n m_i \delta _{x_i}$, 
and, as a result,  
the equation \eqref{eq:lax} will be assumed to hold in the sense of distributions.  
It is helpful to bring 
\eqref{eq:lax1} to its canonical form \cite{ls-cubicstring}. To this 
end one performs a Liouville transformation on \eqref{eq:lax1}.  
This is fully explained in \cite{ls-cubicstring} and here we only state the essential results leading to the appearance of the cubic string boundary 
value problem.   
\begin{lemma} 
Under the change of variables
  \begin{equation}
    \label{eq:liouville}
    \xi =\tanh\frac{x}{2},
    \qquad
    \psi(x)=\frac{2\,\phi(\xi)}{1-\xi^2},
  \end{equation}
  the DP spectral problem \eqref{eq:lax1}
  is equivalent to the cubic string problem
  \begin{equation}
    \label{eq:cubic-spectral}
    \begin{split}
      -\phi_{\xi \xi \xi }(\xi) &= z \, g(\xi) \, \phi(\xi)
      \quad\text{for $\xi \in (-1,1)$},
      \\
      \phi(-1) = \phi_{\xi} (-1) &= 0,
      \\
      \phi(1) &= 0,
    \end{split}
  \end{equation}
  where
  \begin{equation}
    \label{eq:gm}
    \left( \frac{1-\xi^2}{2} \right)^3 g(\xi) = m(x).
  \end{equation}
  In the discrete case,
  when $m(x)=2 \sum_1^n m_i \, \delta_{x_i}$,
  equation \eqref{eq:gm} should be interpreted as
  \begin{equation}
    \label{eq:measure-g}
    g(\xi) = \sum_{i=1}^n g_i \, \delta _{\xi_i},
    \quad\text{where}\quad
    \xi_i = \tanh \frac{x_i}{2},
    \quad
    g_i = \frac{8m_i}{\bigl( 1-\xi_i^2 \bigr)^2}.
  \end{equation}
\end{lemma}
To solve the original DP equation requires solving the spectral and inverse spectral 
problem as explained in detail in \cite{ls-cubicstring} for the case of the finite discrete measure 
$m$.  Our goal in the remainder of the paper is to present a few essential aspects 
of the spectral problem associated to the cubic string for a variety of boundary 
conditions with the view towards explaining the role of a novel class of Cauchy biorthogonal 
polynomials introduced in \cite {Paper1, Paper2}.  These polynomials  can be defined for arbitrary 
positive measures $d\alpha, d\beta$ supported on the 
positive part of the real axis $\mathbb{R}_+$ provided that 
all (Cauchy) bimoments $I_{ij}=\iint \frac{x^iy^j} {x+y} d\alpha(x) d\beta(y) $ are finite.  One then defines polynomials $q_j(y), p_j(x)$ 
of degree $j=0,1, \dots$ satisfying the biorthogonality condition
$\iint \frac{p_j(x)q_k(y)}{x+y} d\alpha(x) d\beta(y) =\delta_{jk}. $ 
We refer to these as Cauchy biorthogonal polynomials.  

\section{Discrete cubic strings}\label{sec:cubic stringb}
We slightly generalize the cubic string discussed in the previous section 
in connection with the DP equation. We recall that in an ordinary string 
problem different boundary conditions correspond to different ways of
tying down the ends of the string. For us, different boundary conditions 
will eventually lead to different spectral measures with respect to 
which we will define biorthogonal polynomials.   However, 
in all cases discussed by us the spectrum is positive and simple, as one would expect
from any vibrating system.   
  
\bd \label{def:csBVP}
{\sl {The cubic string boundary value problems:}}
\begin{equation}
    \label{eq:cubicstring}
    \begin{split}
    &  -\phi_{\xi \xi \xi }(\xi) = z \, g(\xi) \, \phi(\xi), \qquad 
      0<\xi<1, \qquad \phi(0) = \phi_{\xi} (0) = 0,
      \\
     & \text{Type 0 (DP case )}: \phi(1) = 0,\quad \text{Type 1}:\phi_{\xi}(1)=0 \quad \text{Type 2}: \phi_{\xi\xi}(1)=0
    \end{split}
  \end{equation} 
  \ed
  \br For simplicity we have adjusted the length of the string; it is now $1$ rather than $2$.  
  \er 
  We are only interested in the case where 
the mass distribution consists of a finite collection of point-masses: 
  \be  \label{def:gdiscrete}
       g(\xi) = \sum_{i=1}^n g_i \, \delta _{\xi_i}, 
    \quad\text{where}\quad 0< g_i, \quad 0< \xi _1< \xi _2< \dotsb < \xi _n<1.  
  \ee
We will consider all three boundary value problems mentioned above with 
this mass distribution, as well as 
one degenerate case in which the last mass is placed at $1$ 
(i.e. $\xi_n=1$: in the latter case 
we take the right hand limit to compute the derivatives of $f$ at $1$.  
Moreover, for that case, we consider only the BVP of type 2.)  

We will collectively 
refer to all these cases as {\em the discrete cubic string problem}.  

We will also use an accompanying initial value problem,   
\begin{equation}\label{eq:Icubicstring}
\phi (0) = \phi_{\xi} (0) = 0, \phi_{\xi \xi}(0)=1.  
 \end{equation}          
The boundary value problems in Definition \ref{def:csBVP} are  not self-adjoint 
and the adjoint boundary value problems play an important role.  

\bd \label{def:csBVP*}{\sl {The adjoint cubic string boundary value problems:}}
\begin{equation}
    \label{eq:cubicstring*}
         \phi^*_{\xi \xi \xi }(\xi) = z \, g(\xi) \, \phi^*(\xi), \qquad 0<\xi <1,
\end{equation}
Type 0: $\phi^*(1) =\phi^*_{\xi}(1)= 0, \phi^*(0)= 0,$   Type 1: $\phi^*(1) =\phi^*_{\xi \xi}(1)= 0, \phi^*(0)= 0,$ \\Type 2: $\phi^*_{\xi}(1) =\phi^*_{\xi \xi}(1)= 0, \phi^*(0)= 0$.
 \ed
The corresponding initial value problems are: 
\bd \label{def:csIVP*}{\sl {The adjoint cubic string initial value problems:}}
\begin{equation}\label{eq:Icubicstring*}
    \phi^*_{\xi \xi \xi }(\xi) = z \, g(\xi) \, \phi^*(\xi)
      \qquad 0<\xi <1 
 \end{equation} 
with nonzero initial values: Type 0: $\phi^*_{\xi\xi}(1)=1$, Type 1: $\phi^*_{\xi}(1)=1$, Type 2: 
$\phi^*(1)=1.$
\ed
Since all three boundary value problems for $\phi$ satisfy the same initial value problem we will use one letter, namely $\phi$, to denote the solution.  However, we will attach an index $a=0,1,2$ to $\phi^*$ 
to indicate the type of the BVP; for example, $\phi^{*}_0$ will refer to the BVP/IVP  of Type $0$  etc.  

In the process of integration by parts of expressions like $\int_0^1 f_{\xi \xi \xi}(\xi) h(\xi) d\xi$ one identifies the relevant 
bilinear symmetric form: 
\bd Given any twice differentiable $f,h$ the bilinear concomitant is 
defined as the bilinear form:
\be
B(f,h)(\xi) =f_{\xi \xi}h-f_{\xi}h_{\xi}+fh_{\xi \xi}.  
\ee
\ed
This bilinear symmetric form induces a bilinear symmetric form (denoted also by B) on triples $F^T:=(f,f_{\xi}, f_{\xi \xi})$, namely 
\be \label{def:J}
B(F,H)=F^TJH, \quad J:=\begin{pmatrix}0&0&1\\
                                                     0&-1&0\\
                                                      1&0&0 \end{pmatrix}.  
\ee
 Furthermore, we 
also define a natural $L^2$ space associated with $g$, denoted $L^2[0,1]_g$, 
equipped with the inner product: $
(f,h)_g=\int _0^1 f(\xi)h(\xi) g(\xi) \, d\xi$.  Whenever $f$ or $g$ depend on the spectral variables $z$ and $ \lambda$, we write $(f(z),g(\lambda))_g$ to display this dependence.  

Since all initial value problems \ref{eq:Icubicstring}, \ref{def:csIVP*} can be solved 
for arbitrary $z\in \C$, $\phi$ and $\phi_a^*$ are functions of the spectral parameter $z$.    The following theorem establishes a relation between these two 
functions in terms of their concomitant and the relevant boundary value problem.   The customary notation: $f(\xi) |_0^1=f(1)-f(0)$ is used in the statement below.  
\bl 
Suppose $\phi(\xi;z)$ and $\phi_a^*(\xi; \lambda)$ are 
solutions to the IVPs \ref{eq:Icubicstring}, 
\ref{def:csIVP*} with  
spectral parameters $z$, $\lambda$ respectively.  Then 
\begin{enumerate}
\item Type 0: the spectrum is determined by the zeros of $\phi(1,z)=0$ and 
\be
-B(\phi(\xi;z), \phi_0^*(\xi, \lambda))|_0^1=\phi_0^*(0;\lambda)-\phi(1;z)=(z-\lambda)
(\phi(z), \phi_0^*(\lambda))_g. \label{eq:0fidentity}
 \ee
\item Type 1: the spectrum is determined by the zeros of $\phi_{\xi} (1,z)=0$ and 
\be
-B(\phi(\xi;z), \phi_1^*(\xi, \lambda))|_0^1=\phi_1^*(0;\lambda)+\phi_{\xi}(1;z)=(z-\lambda)
(\phi(z), \phi_1^*(\lambda))_g. \label{eq:1fidentity}
\ee
\item Type 2: the spectrum is determined by the zeros of $\phi_{\xi \xi}(1,z)=0$ and 
\be
-B(\phi(\xi;z), \phi_2^*(\xi, \lambda))|_0^1=\phi_2^*(0;\lambda)-\phi_{\xi \xi}(1;z)=(z-\lambda)
(\phi(z), \phi_2^*(\lambda))_g.  \label{eq:2fidentity}
 \ee
\end{enumerate}
In addition, 
\be \label{eq:ffidentity}
-B(\phi(\xi;z),\phi(\xi;\lambda))|_0^1=(z+\lambda)(\phi(z),\phi(\lambda))_g.  
\ee

\el
\begin{proof}
Indeed \eqref{eq:Icubicstring} and two integrations by parts imply that 
\bew
\begin{split}
-\int_0^1 \phi_{\xi \xi \xi}(\xi;z)\phi_a^*(\xi;\lambda)\, d\xi =&-B(\phi, \phi_a^*)|_0^1+\int_0^1 \phi(\xi;z)\phi^*_{a,\xi \xi \xi}(\xi;\lambda)\, d\xi =\\
&z\int_0^1 \phi(\xi;z)
\phi_a^*(\xi;\lambda)g(\xi) \, d\xi. 
\end{split}
\eew 
Consequently, using equation \eqref{eq:Icubicstring*} we obtain: 
\bew
-B(\phi, \phi_a^*)|_0^1=(z-\lambda)\int_0^1 \phi(\xi;z)
\phi_a^*(\xi;\lambda)g(\xi) \, d\xi,   
\eew
which in view of the initial conditions implies the claim.  
A similar computation works for the second identity.  
\end{proof}
By specializing $\lambda=z$ in the lemma above one readily obtains: 
\bc \label{cor:basicinstinct}
$\phi$ and $\phi_a^*$ satisfy the following relations: 
\begin{enumerate}
\item Case 0: $\phi_0^*(0;z)=\phi(1;z)$. Case 1: $\phi_1^*(0;z)=-\phi_{\xi}(1;z)$, Case 2: $\phi_2^*(0;z)=\phi_{\xi\xi}(1;z)$.  
\item Case 0: $-\phi_{z}(1;z)=(\phi(z),\phi_0^*(z))_g$.
Case 1: $\phi_{\xi z}(1;z)=(\phi(z),\phi_1^*(z))_g$.
Case 2: $-\phi_{\xi \xi z}(1;z)=(\phi(z),\phi_2^*(z))_g$.  
\end{enumerate}
\ec 
We give below a complete characterization of the spectra and 
the corresponding eigenfunctions  
for all three BVPs.  We also select certain combinations of BVPs 
which reveal the origin of the relevance of the Cauchy kernel $\frac{1}{x+y}$ 
to the spectral theory of the cubic string.  

\bt \label{thm:spectrum}
Consider a cubic string with a finite measure $g$ as in 
\eqref{def:gdiscrete}.      
\begin{enumerate}
\item Let $z_{a,j}$ denote the eigenvalues of the BVP of Type $a=0,1,2$. 
In each of the three cases, the spectrum is positive and simple.   
\item For any pair of BVPs of Type 
0,1,2 the spectra are interlaced in the following order: 
$$
z_{2,j}<z_{1,j}<z_{0,j},   \quad j=1,\dots n
$$
\item 
For each $a=0,1,2$ the eigenfunctions $\phi(\xi;z_{a,j})\equiv \phi_{a,j}(\xi)$ 
are linearly independent.  

\item For the following
combinations of BVPs, 
$(\phi_{a,i},\phi_{b,j})_g$ factorizes: 
\begin{enumerate}
\item Type 00: 
\be 
(\phi_{0,i},\phi_{0,j})_g=\frac{\phi_{0,i,\xi}(1)\phi_{0,j,\xi}(1)}{z_{0,i}+
z_{0,j} }
\ee
\item Type 01
\be 
(\phi_{0,i},\phi_{1,j})_g=-\frac{\phi_{0,i,\xi\xi}(1)\phi_{1,j}(1)}{z_{0,i}+
z_{1,j}} 
\ee
\item Type 12 
\be 
(\phi_{1,i},\phi_{2,j})_g=-\frac{\phi_{1,i,\xi\xi}(1)\phi_{2,j}(1)}{z_{1,i}+
z_{2,j}} 
\ee
\item Type 22
\be 
(\phi_{2,i},\phi_{2,j})_g=\frac{\phi_{2,i,\xi}(1)\phi_{2,j,\xi}(1)}{z_{2,i}+
z_{2,j} }
\ee
\end{enumerate}
 
\end{enumerate}
\et

\begin{proof}
It is easy to check (see Section 4.1 in \cite{ls-cubicstring}) that 
\be \label{eq:matrixphi}
\begin{pmatrix}
    \phi(1;z) \\
    \phi_{\xi}(1;z) \\
    \phi_{\xi \xi }(1;z)
  \end{pmatrix}
  =
  L_{n}G_n(z) \, L_{n-1} \, G_{n-1}(z) \cdots L_1 \, G_1(z) \, L_0
  \begin{pmatrix}
    0 \\ 0 \\ 1
  \end{pmatrix}.
\ee
where 
\bew 
G_k(z)=\begin{pmatrix}
    1 & 0 & 0 \\
    0 & 1 & 0 \\
    -z\,g_k & 0 & 1
  \end{pmatrix}, 
\eew
\bew
L_k =
  \begin{pmatrix}
    1 & l_k & l_k^2/2 \\
    0 & 1 & l_k \\
    0 & 0 & 1
  \end{pmatrix}, 
\eew
and
\begin{equation}
  \label{eq:lk}
  l_k = \xi _{k+1} - \xi _k, \qquad  \xi_0=0, \qquad  \xi_{n+1}=1.  
\end{equation}
We prove the statement about the spectra by using the results obtained 
in \cite{ls-cubicstring}.  
By Theorem 3.5 in 
\cite{ls-cubicstring}  $\phi(1;z)$ has $n$ distinct positive zeros and 
so do $\phi_{\xi}(1;z)$ and $\phi_{\xi \xi}(1;z) $ (denoted there 
$\phi_{y}, \phi_{yy}$). The second statement follows if one observes that 
$\frac{\phi_{\xi}(1;z)}{\phi_{z}(1;z)}$ and  
$\frac{\phi_{\xi\xi}(1;z)}{\phi_{z}(1;z)}$ are strictly positive on the spectrum of 
Type 0 being the residues of $\frac{\phi_{\xi}(1;z)}{\phi_(1;z)}$, $\frac{\phi_{\xi \xi }(1;z)}{\phi_(1;z)}$ respectively (Theorem 3.5 and Theorem 3.15 in \cite{ls-cubicstring}).  It follows then that both $\phi_{\xi}(1;z)$ and $\phi_{\xi \xi}(1;z)$ 
change signs $n$ times, hence all three spectra are simple.  Furthermore, 
$\phi_{\xi}(1;z)>0$, $\phi_{\xi\xi}(1;z)>0$ for $z\leq 0$, so the zeros of 
$\phi_{\xi}(1;z)$ and $\phi_{\xi \xi}(1;z)$ are strictly positive.  Since $\phi_{\xi}(1;z)$ and $\phi_{\xi \xi}(1;z)$ change sign on every interval between two consecutive zeros of $\phi(1;z)$ the spectra of Type 0 and 1, as well 
as 0 and 2, interlace. 
To see that the spectrum of Type 1 interlaces with the spectrum of Type 2 
we proceed as follows.  By \eqref{eq:ffidentity}, after evaluating at 
$z=z_{2,i}, \lambda=-z_{2,i}$, we obtain
\bew 
\phi_{\xi}(1;z_{2,i})\phi_{\xi}(1;-z_{2,i})=\phi_{\xi \xi} (1; -z_{2,_i})
\phi(1;z_{2,i}), 
\eew
which gives
\be \label{eq:alt}
\phi_{\xi}(1;z_{2,i})=\frac{\phi_{\xi \xi} (1; -z_{2,i})}
{\phi_{\xi}(1;-z_{2,i})}\phi(1;z_{2,i})
\ee
For $z>0$, $\sgn(\phi_{\xi}(1;-z))=\sgn(\phi_{\xi \xi}(1;-z))=+1$ because 
both are strictly positive there.  
Since the zeros of $\phi(1;z)$ interlace with the zeros of 
$\phi_{\xi \xi}(1;z)$, $\sgn(\phi(1;z_{2,i}))$ alternates, which in turn 
implies that $\sgn(\phi_{\xi}(1;z_{2,i}))$ alternates as well.  Thus the zeros of $\phi_{\xi} (1;z)$
interlace with the zeros of $\phi_{\xi \xi}(1;z)$.  

The relative position of the spectra of the three types is 
best inferred from the fact that on the first eigenvalue $z_{0,1}$ of Type 0, $\phi_{\xi}(1;z_{0,1})$ and 
$\phi_{\xi \xi}(1;z_{0,1})$ are both negative since $\frac{\phi_{\xi}(1,z)}{\phi_{z}(1,z)}$ and  
$\frac{\phi_{\xi\xi}(1,z)}{\phi_{z}(1,z)}$ are strictly positive on the spectrum of 
Type 0.  Thus the spectra of Type 1 and 2 are shifted to the left relative 
to the spectrum of Type 0. In particular, $\phi(1,z_{2,1})>0$ and so is $\phi_{\xi}(1,z_{2,1})$
by \eqref{eq:alt}.  Thus, at least the first zero of $\phi_{\xi \xi}$ occurs 
to the left of the zeros of $\phi_{\xi}$ and $\phi$. So $ z_{2,1}<z_{1,1}<z_{0,1}$. Suppose this holds for the $(j-1)$st eigenvalues.  Then we know that 
both $z_{0,j-1}<z_{1,j}< z_{0,j}$ and $z_{0,j-1}<z_{2,j}<z_{0,j}$.  If 
$z_{1,j}<z_{2,j}$ then $z_{2,j-1}<z_{1,j-1}<z_{1,j}<z_{2,j}$, thus contradicting that the spectra of type 1 and 2 interlace.    

As for the linear independence we observe that the cubic string boundary value problem \ref{def:csBVP} can be  
written as an integral equation:
\be 
\phi(\xi;z) =z \int_0^1 G(\xi, \tau)\phi(\tau;z)g(\tau)\, d\tau , 
\ee
 where $G(\xi, \tau)$ is the Green function satisfying the 
boundary conditions of \ref{def:csBVP}.  
Then the linear independence of eigenfunctions corresponding to 
distinct eigenvalues follows from the general result about the 
eigenfunctions of a linear operator.  

Finally, the statements about the eigenfunctions follow immediately from equation
\eqref{eq:ffidentity} after setting $z=z_{a,i}, \lambda=z_{b,j}$. 
 
\end{proof}
We now briefly analyze the degenerate case with the mass $m_n$ at the 
end point $x_n=1$.  We do it only to illustrate that even though the spectrum 
degenerates in this case the overall conclusions hold.  

\bt \label{thm:spectrum2}
Let us consider a cubic string with a finite measure $g$ as in 
\eqref{def:gdiscrete} with $x_n=1$.  Then  
\begin{enumerate}
\item the spectra of Type 0, 1, and 2 are positive, simple and their cardinalities are $n-1$
for Type 0 and Type 1 and 
$n$  for Type 2,
\item the spectra interlace 
\bew 
  0<z_{2,1}<z_{1,1}<z_{0,1} <\dotsb <z_{2,n-1}<z_{1,n-1}< z_{0,n-1} < z_{2,n},
\eew  
\item   the eigenfunctions $\phi(\xi;z_{2,j}):= \phi_{2,j}(\xi)$ satisfy 
\be 
(\phi_{2,i},\phi_{2,j})_g=\frac{\phi_{2,i,\xi}(1)\phi_{2,j,\xi}(1)}{z_{2,i}+z_{2,j}}, 
\ee
and they are linearly independent. 
\end{enumerate}
\et

\br We notice that $\deg \phi(1;z)=\deg \phi_{\xi}(1;z)=n-1$ while 
$\deg \phi_{\xi\xi}(1;z)=n$.  This is in contrast to the previous cases 
with all positions $x_1, \dots, x_n$ inside the interval $[0,1]$ for which 
all polynomials have the same degree $n$.  
\er 
\begin{proof}
The spectrum of Type 2 is clearly given by the zeros of $\phi_{\xi \xi}(1;z)$. 
 Let us first consider the case when $m_n$ is placed slightly to the left of the 
point $1$.  Thus, initially, $l_n >0$ (see \eqref{eq:matrixphi}).  
By Theorem \ref{thm:spectrum} $\phi(1;z)$ has $n$ distinct positive zeros and 
so does $\phi_{\xi \xi}$ and they interlace.   We subsequently 
take the limit $l_n\rightarrow 0$ in the above formulas. We will 
use the same letters for the limits to ease the notation. 
   By simple 
perturbation argument, $z_{n,0} \rightarrow \infty$.  Since $z=0$ is not in the spectrum, $z_{2,1}$ has to stay away from $0$.  This shows that the spectrum is positive. Furthermore, in the limit $z_{0,1}, \dots, z_{0,n-1}$ approach simple zeros of the BVP of type 0 for $n-1$ masses. Indeed, using \eqref{eq:matrixphi} with $l_n=0$ there, we obtain: 
\be \label{eq:lasteq}
\phi(1+0;z)=\phi(1-0;z), \quad \phi_{\xi}(1+0;z)=\phi_{\xi}(1-0;z), \quad 
\phi_{\xi \xi}(1+0;z)=-zm_n \phi(1-0;z)+\phi_{\xi \xi}(1-0;z), 
\ee 
where $1\pm0$ refers to the right hand, the left hand limit at $1$ respectively.  
To see that the spectrum is simple
we observe that if in the limit two successive eigenvalues coalesce, namely 
$z_{2,i}=z_{2,i+1}$, then necessarily $z_{2,i}=z_{0,i}$ because 
of the interlacing property.  However now equation \eqref{eq:lasteq} implies 
that $\phi(1-0;z_{0,i})=\phi_{\xi \xi}(1-0;z_{0,i})=0$ which contradicts 
Theorem \ref{thm:spectrum} for the BVP of Type 0 for $n-1$ masses.  Thus, 
the zeros of $\phi(1,z)$ and $\phi_{\xi \xi}
(1+0,z)$ interlace and we have 
\bew 
  0<z_{2,1}<z_{0,1}<\dotsb <z_{2,n-1}<z_{0,n-1} < z_{2,n}.  
\eew 
Likewise, for the spectrum of Type 1, $z_{n,1}\rightarrow \infty$ and the 
remaining roots interlace according to the pattern valid for $n-1$ masses.  

To prove the statement about the eigenfunctions we use 
\eqref{eq:ffidentity} and after setting $z=z_i, \lambda=z_j$ 
in that formula 
we obtain the required identity.  
The linear independence is proven by the same type of argument as in the 
proof of Theorem \ref{thm:spectrum}.  
\end{proof}
We immediately have several results about the adjoint cubic string 
\ref{def:csBVP*}.  

\bc 
Given a discrete finite measure $g$
\begin{enumerate}
\item for each of the three types of the BVPs the 
adjoint cubic string (Definition \ref{def:csBVP*}) and the cubic string 
(Definition \ref{def:csBVP}) have identical spectra.
\item  the families of eigenfunctions $\{\phi_{a,j}\}$ 
and $\{\phi_{a,j}^* \}$ are 
biorthogonal, that is: 
\be 
(\phi_{a,i},\phi_{a,j}^*)_g=0 \, \qquad \text{whenever}\qquad  i\neq j. 
\ee 
\item For $i=j$, 
\be \label{eq:diagpair}
(\phi_{a,i},\phi_{a,i}^*)_g=\begin{cases}
-\phi_{z}(1;z_{0,i})\neq 0, \qquad &a=0\\
\phi_{\xi z}(1;z_{1,i})\neq 0, \qquad &a=1\\
-\phi_{\xi \xi z}(1;z_{2,i})\neq 0, \qquad &a=2 \end{cases}
\ee
\end{enumerate}
holds.  
\ec 
\begin{proof}
The first equality in Corollary \ref{cor:basicinstinct} implies that the 
spectra of the cubic string and its adjoint are identical.  
The biorthogonality follows immediately from equations \eqref{eq:0fidentity}, \eqref{eq:1fidentity} and \eqref{eq:2fidentity}.  
For $i=j$ we use Corollary \ref{cor:basicinstinct}.  Finally, 
since the spectrum is simple the required derivatives with respect to $z$ are 
nonzero.   
\end{proof}
 We conclude this section with the definition and some fundamental properties of two important functions which play a significant role in the theory (Section 6 in \cite{Paper1}).  
 \bd \label{def:Weylfs}
 The following functions are called Weyl functions for their respective BVPs:
 \begin{align*}
 &\text{Type 0 }: W(z):=\frac{\phi_{\xi}(1;z)}{\phi(1;z)}, \quad Z(z)=\frac{\phi_{\xi\xi}(1;z)}{\phi(1;z)}; \qquad 
 \text{Type 1 }:  W(z):=-\frac{\phi(1;z)}{\phi_{\xi}(1;z)}, \quad Z(z)=\frac{\phi_{\xi\xi}(1;z)}{\phi_{\xi}(1;z)}; \\
&\text{Type 2 }: W(z):=-\frac{\phi_{\xi}(1;z)}{\phi_{\xi \xi}(1;z)}, \quad Z(z)=-\frac{\phi(1;z)}{\phi_{\xi \xi}(1;z)}.
 \end{align*}
  \ed
 \br The definition of the Weyl functions for the BVP of type 2 in the 
degenerate case is identical to the one given above for the BVP of type 2.  \er

The Weyl functions $W$ and $Z$ are not independent, they are related by an identity 
 which was originally formulated for the DP peakons in \cite{ls-cubicstring}. 
As an example we formulate such an identity for the BVP of type 2 
(both the degenerate as well as the nondegenerate case).  
\bl \label{lem:WZidentity}
Consider the BVP of type 2.  Then the corresponding Weyl functions satisfy:
\be 
W(z)W(-z)+Z(z)+Z(-z)=0
\ee
\el 
\begin{proof}
By formula \eqref{eq:ffidentity}
\be 
B(\phi(\xi;z), \phi(\xi;-z))|_0^1=0,  
\ee
which, when written out explicitly, 
gives the identity: 
\be 
\phi_{\xi\xi}(1;z)\phi(1;-z)-\phi_{\xi}(1;z)\phi_{\xi}(1;-z)+
\phi_{\xi\xi}(1;-z)\phi(1;z)=0.  
\ee
Upon dividing the last equation by $\phi_{\xi\xi}(1;z)\phi_{\xi \xi}(1;-z)$ 
we obtain the claim. 
\end{proof}

 We state now the spectral representation theorem for $W(z)$ and $Z(z)$ 
 for the BVP of Type 2 in the degenerate case, the 
remaining cases being merely variations of this, most transparent case. 
One recognizes again the presence of the Cauchy kernel $\frac{1}{x+y}$ in the 
spectral representation of $Z(z)$.   
 \bt \label{thm:WZprop}
Consider the BVP of type 2 (degenerate case).  Then 
 the Weyl functions $W$ and $Z$ have the following (Stieltjes) integral 
representations: 
 \be 
 W(z)=\int \frac{1}{z-y}d\beta(y), \qquad Z(z)=\int \frac{1}{(z-y)(x+y)}d\beta(y) d\beta(x), 
 \ee
 where $d\beta =\sum_{i=1}^n b_i \delta_{z_{2,i}}$ and $b_i=\frac{-\phi_{\xi}(1;z_{2,i})}{\phi_{\xi \xi z}(1;z_{2,i})}>0$.  
 \et
 \begin{proof}
Since $\phi_{\xi}(1;z), \phi_{\xi \xi}(1;z) $ have simple, interlacing zeros, 
and $\deg \phi_{\xi} (1;z) =n-1$ while $\deg \phi_{\xi \xi}(1;z)=n$, 
$W(z)$ admits a partial fraction decomposition with simple factors:
\bew
W(z)=\sum_{i=1}^n \frac{b_i}{z-z_{2,i}}, 
\eew
where, by the residue calculus, 
$b_i=\frac{-\phi_{\xi}(1;z_{2,i})}{\phi_{\xi \xi z}(1;z_{2,i})}$.  Moreover the $b_i$s  are all of the same sign because the zeros 
of $\phi_{\xi} (1;z)$ and $\phi_{\xi \xi}(1;z)$ interlace and, consequently, 
it suffices to check the sign of $\frac{-\phi_{\xi}(1;z_{2,i})}{\phi_{\xi \xi z}(1;z_{2,i})}$ at the first zero $z_{2,1}$. 
By Theorem \ref{thm:spectrum} 
$\sgn(\phi_{\xi}(1;z_{2,1})=1$, and thus $b_1>0$ because on the first zero 
$\phi_{\xi \xi z}$ must be negative.  Consequently, all $b_i>0$.   

Likewise, $Z(z)$ 
admits a partial fraction decomposition: 
\bew
Z(z)=\sum_{i=1}^n \frac{c_i}{z-z_{2,i}}, 
\eew
and again, it follows from the second item in Theorem \ref{thm:spectrum}
that $c_i>0$.  Finally, by residue calculus, it follows from Lemma
\ref{lem:WZidentity} that 
\bew
c_i=\sum_{j=1}^n \frac{b_i b_j}{z_{2,i}+z_{2,j}}, 
\eew
which proves the integral representation for $Z(z)$.  

\end{proof}
\section{Generalized Fourier transform and biorthogonality} \label{sec:cubicstringc}
Since $\phi_{a,i}$ are linearly independent we can decompose any $f\in L_g^2[0,1]$ 
in the basis of $\{\phi_{a,i}\}$ and use the dual family $\{\phi_{a,i}^*\}$ 
to compute the coefficients in the expansion: 
$$
f=\sum_i C_{a,i} \phi_{a,i}, \quad C_{a,i}=\frac{(\phi_{a,i}^*,f)_g}
{(\phi_{a,i}^*,\phi_{a,i})_g}.  
$$
For each pair $a, b$ for which $(\phi_{a,i},\phi_{b,j})_g$ factorizes (item 
3 in Theorem \ref{thm:spectrum})
we define two (finite dimensional) Hilbert spaces $H_{\alpha}:=L^2({\R, d\alpha})$ and $H_{\beta}
:=L^2({\R, d\beta}) $ where the measures $d\alpha$ and $d\beta$ are 
chosen as follows: 1) using Theorem \ref{thm:spectrum} split the numerator of $(\phi_{a,i},\phi_{b,j})_g$ , 2) perform the partial fraction decomposition of the Weyl functions with numerators matching the factors in $(\phi_{a,i},\phi_{b,j})_g$ , 3) represent 
the partial fraction decompositions as Stieltjes' transforms of the respective measures. 

\bx 
For Type 00, item 3 in Theorem \ref{thm:spectrum} states: 
$
(\phi_{0,i},\phi_{0,j})_g=\frac{\phi_{0,i,\xi}(1)\phi_{0,j,\xi}(1)}{z_{0,i}+
z_{0,j} }
. $
Hence the numerator splits into $\phi_{0,i,\xi}(1)$ and $\phi_{0,j,\xi}(1)$ and the corresponding 
Weyl function will be $W(z)=\frac{\phi_{\xi}(1;z)}{\phi(1;z)}$ (taken twice) with partial fraction decomposition 
$W(z)=\int \frac{1}{z-x}d\alpha(x)$ where $\d\alpha=d\beta=\sum_{i} \frac{\phi_{0,i.\xi}}{\phi_{z}(1;z_{0,i})}\delta_{z_{0,i}}$
For Type 01, item 3 in Theorem \ref{thm:spectrum} states $
(\phi_{0,i},\phi_{1,j})_g=-\frac{\phi_{0,i,\xi\xi}(1)\phi_{1,j}(1)}{z_{0,i}+
z_{1,j}} $ and the numerator splits into $\phi_{0,i,\xi\xi}(1)$ and $\phi_{1,j}(1)$ which, in turn, 
match with $Z(z)$ for Type 1 and $W(z)$ for Type 2 in the Definition \ref{def:Weylfs} .  The resulting partial fractions 
decompositions $Z(z)=\int \frac{1}{z-x}d\alpha(x), \, W(z)=\int \frac{1}{z-y}d\beta(y)$
imply 
$
\d\alpha=\sum_{i} \frac{\phi_{0,i,\xi\xi}}{\phi_{z}(1;z_{0,i})}\delta_{z_{0,i}}, 
\d\beta =-\sum_{i} \frac{\phi_{1,i}}{\phi_{\xi z}(1;z_{1,i})}\delta_{z_{1,i}}. 
$

\ex

In summary, for every pair $\{a,b\}$ of BVPs for the cubic string we 
associate two Hilbert spaces $H_{\alpha}$ and $H_{\beta}$ with the pairing
$H_{\alpha}\times H_{\beta}\rightarrow \C$: 
\bd 
$$
\langle p|q\rangle=\int \frac{p(x)q(y)}{x+y}d\alpha(x) \beta(y), \quad 
 p\in H_{\alpha}, \quad q \in H_{\beta}. 
$$
\ed
We now introduce a family of generalized Fourier transforms adapted to 
each of the three types of BVPs 
\bd 
Given $f\in L_g^2[0,1]$ and $a=0,1,2$ 
\be \label{def:spectralmap} 
\hat f_a(z):=(-1)^a \int_0^1 \phi_a^*(\xi;z)f(\xi) g(\xi)d\xi.  
\ee
\ed 
\br 
\label{rem:GF}
Observe that 
$
\hat f_a(z)=((-1)^a\phi_a^*(z),f)_g, 
$
and, in particular, 
$
\hat f_a(z_{a,i})=((-1)^a\phi^*_{a,i},f)_g
$
whenever $z$ equals to one of the points of the spectrum of \eqref{eq:cubicstring}.  
\er

\br A map of this type was introduced in the context of the inhomogeneous string 
problem by I.S. Kac and M.G. Krein in \cite{kackrein} as a generalization of the Fourier transform.  A similar map known as a distorted Fourier transform is commonly 
used in Quantum Mechanics.  
\er
The main property of this map is captured in the following theorem. 
\bt \label{thm:isometry}
For every pair $a, b$ for which $(\phi_{a,i},\phi_{b,j})_g$ factorizes
 (item 3 in Theorem \ref{thm:spectrum})
the generalized Fourier transforms satisfy Parseval's identity, that is,  
for every $f,h \in L_g^2[0,1]$ 
\be
 (f,h)_g=\langle\hat f_a|\hat h_b\rangle=\langle\hat h_a|\hat f_b\rangle.   
\ee

\et 
\begin{proof}
Let us fix $a,b$ for which $(\phi_{a,i},\phi_{b,j})_g$ factorizes.  
Consider two functions $f,h\in L_g^2[0,1]$.  Writing their expansions in the 
bases $\{\phi_{a,i}\}$, $\{\phi_{b,i}\}$ respectively, we obtain 
\bew 
f=\sum_i \frac{(\phi^*_{a,i},f)_g}{(\phi_{a,i}^*,\phi_{a,i})_g} \phi_{a,i}, \quad 
h=\sum_j \frac{(\phi^*_{b,j},h)_g}{(\phi_{b,j}^*,\phi_{b,j})_g} \phi_{b,j}.  
\eew
Hence their inner product reads: 
\bew
(f,h)_g=\sum_{i,j}\frac{(\phi^*_{a,i},f)_g}{(\phi_{a,i}^*,\phi_{a,i})_g}\frac{(\phi^*_{b,j},h)_g}{(\phi_{b,j}^*,\phi_{b,j})_g}(\phi_{a,i},\phi_{b,j})_g
\eew
Applying now item $3$ from Theorem \ref{thm:spectrum} as well as 
item $2$ from Lemma \ref{cor:basicinstinct} we obtain 
\begin{align*}
Type\, 00: &(f,h)_g=\sum_{i,j}\frac{(\phi^*_{0,i},f)_g}{\phi_{z}(1;z_{0,i})}\frac{(\phi^*_{0,j},h)_g}{\phi_{z}(1;z_{0,j})}\frac{\phi_{\xi}(1;z_{0,i})\phi_{\xi}(1;z_{0,j})}{z_{0,i}+z_{0,j}},\\
Type\, 01: &(f,h)_g=\sum_{i,j}\frac{(\phi^*_{0,i},f)_g}{\phi_{z}(1;z_{0,i})}\frac{(\phi^*_{1,j},h)_g}{\phi_{\xi z}(1;z_{1,j})}\frac{\phi_{\xi\xi}(1;z_{0,i})\phi(1;z_{1,j})}{z_{0,i}+z_{1,j}},\\ 
Type\, 12: &(f,h)_g=\sum_{i,j}\frac{(\phi^*_{1,i},f)_g}{\phi_{\xi z}(1;z_{1,i})}\frac{(\phi^*_{2,j},h)_g}{\phi_{\xi \xi z}(1;z_{2,j})}\frac{\phi_{\xi\xi}(1;z_{1,i})\phi(1;z_{2,j})}{z_{1,i}+z_{2,j}},\\
Type\, 22: &(f,h)_g=\sum_{i,j}\frac{(\phi^*_{2,i},f)_g}{\phi_{\xi \xi z}(1;z_{2,i})}\frac{(\phi^*_{2,j},h)_g}{\phi_{\xi \xi z}(1;z_{2,j})}\frac{\phi_{\xi}(1;z_{2,i})\phi_{\xi}(1;z_{j,2})}{z_{2,i}+z_{2,j}}.
\end{align*}
We now define the weights $b_j, a_j$ generating  
the measures $\d\beta=\sum _j B_j\delta _{z_j}$, $d\alpha=\sum _j A_j\delta _{z_j} $ respectively, 
 as residues of $W$s or $Z$s:
\begin{align*}
Type\, 00: A_i&=\frac{\phi_{\xi}(1;z_{0,i})}{\phi_{z}(1;z_{0,i})},\qquad  
B_j=\frac{\phi_{\xi}(1;z_{0,j})}{\phi_{z}(1;z_{0,j})},\\
Type\, 01: A_i&=\frac{\phi_{\xi\xi}(1;z_{0,i})}{\phi_{z}(1,z_{0,i})}, \qquad 
B_j=-\frac{\phi(1;z_{1,j})}{\phi_{\xi z}(1;z_{1,j})},\\
Type\,  12: A_i&=\frac{\phi_{\xi\xi}(1,z_{1,i})}{\phi_{\xi z}(1,z_{1,i})},
\qquad B_j=-\frac{\phi(1,z_{2,j})}{\phi_{\xi\xi z}(1,z_{2,j})},\\
Type \, 22: A_i&=-\frac{\phi_{\xi}(1,z_{2,i})}{\phi_{\xi \xi z}(1,z_{2,i})},
\qquad B_j=-\frac{\phi_{\xi}(1,z_{2,j})}{\phi_{\xi \xi z}(1,z_{2,j})},
\end{align*}
and thus indeed
\bew
(f,h)_g=\sum_{i,j}((-1)^a \phi^*_{a,i},f)_g((-1)^b\phi^*_{b,j},h)_g\frac{A_iB_j}{z_{a,i}+z_{b,j}}.  
\eew
Thus, in view of Remark \ref{rem:GF}, we 
obtain
\bew
(f,h)_g=\sum_{i,j}\hat f_1 (z_{a,i}) \hat h_b(z_{b,j}) \frac{A_iB_j}{z_{a,i}+z_{2,j}}=\int \frac{\hat f_a(x)\hat h_b(y)}
{x+y} d\alpha(x) d\beta(y)=\langle\hat f_a|\hat h_b\rangle.  
\eew
\end{proof}
\br Expanding an arbitrary $f\in L^2_g[0,1]$ 
\bew 
f=\sum_i \frac{(\phi^*_{a,i},f)_g}{(\phi_{a,i}^*,\phi_{a,i})_g} \phi_{a,i}, \quad 
\eew 
allows one to conclude that 
\be \label{eq:Diracdel}
\delta(\xi,\xi'):=\sum _i \frac{\phi_{a,i}(\xi) \phi^*_{a,i}(\xi ')}{(\phi_{a,i}^*,\phi _{a,i})_g}
\ee
plays the role of the Dirac delta on $L^2_g[0,1]$.  
\er
It is now elementary to find the inverse 
Fourier transforms 
\bl
Consider the BVP of type a.  Let $\{z_{a,i}\}$ be the corresponding 
spectrum and let $d\nu _a=\sum_i \delta _{z_{a,i}}$ be an associated 
measure.  Then the inverse generalized Fourier transform  of type a is 
given by 
\be 
(-1)^a \int \hat f_a(z) \frac{\phi(\xi; z)}{(\phi_a^*(z),\phi(z))_g} d\nu _a (z)
\ee
\el 
\begin{proof}
This is a direct computation: 
\bew 
\begin{split}
&(-1)^a \int \hat f_a(z) \frac{\phi(\xi; z)}{(\phi_a^*(z),\phi(z))_g}d\nu_a(z)=
(-1)^a\sum _i \hat f_a(z_{a,i})\frac{\phi_{a,i}(\xi)}{(\phi_{a,i}^*,\phi_{a,i})_g}=
\sum _i (\phi _{a,i}^*, f)_g\frac{\phi_{a,i}(\xi)}{(\phi_{a,i}^*,\phi_{a,i})_g}=\\
&\int_0^1\sum _i \frac{\phi_{a,i}(\xi)}{(\phi_{a,i}^*,\phi _{a,i})_g} \phi^*_{a,i}
(\xi') f(\xi') g(\xi ') d\xi '=
 \int_0^1 \delta(\xi, \xi ') f(\xi')g(\xi ') d\xi'=f(\xi).  
\end{split}
\eew
\end{proof}
There are in general two measures associated with each type of the BVP, 
one generated by $W$ the other by $Z$.  
One can use either one of the them instead of the measure $d\nu$.  We give 
as an example the relevant statement for the the case of the BVP of Type 2, 
both the degenerate and the nondegenerate one.  
\bl The inverse generalized Fourier transform of Type 2 is  
given by 
\be 
\int \hat f_2(z) \frac{\phi(\xi; z)}{\phi_{\xi}(1;z)} d\beta(z)
\ee
\el 
\begin{proof}
~From the definition of $d\beta$ given in Theorem \ref{thm:WZprop} we 
see that 
\bew 
\begin{split}
&\int \hat f_2(z) \frac{\phi(\xi; z)}{\phi_{\xi}(1;z)} d\beta(z)=
\sum _i \hat f_2(z_{2,i})\frac{\phi_{2,i}(\xi)}{\phi_{\xi}(1;z_{2,i})}b_i=
\sum _i \hat f_2(z_{2,i}) \frac{\phi_{2,i}(\xi)}{\phi_{\xi}(1;z_{2,i})}(-\frac{\phi_{\xi}(1;z_{2,i})}{\phi_{\xi \xi z}(1,z_{2,i})})=\\
&\sum _i (\phi ^*_{2,i}, f)_g\frac{\phi_{2,i}(\xi)}{\phi_{\xi}(1;z_{2,i})}(-\frac{\phi_{\xi}(1;z_{2,i})}{\phi_{\xi \xi z}(1;z_{2,i})}=
\sum _i (\phi ^*_{2,i}, f)_g\frac{\phi_{2,i}(\xi)}{(\phi_{2,i}^*,\phi _{2,i})_g},
\end{split}
\eew
where in the last two steps we used Remark \ref{rem:GF} and equation 
\eqref{eq:diagpair} respectively. Thus 
\bew
\int \hat f(z) \frac{\phi(\xi; z)}{\phi_{\xi}(1;z)} d\beta(z)=\int_0^1 
\sum _i \frac{\phi_{2,i}(\xi)\phi_{2,i}^*(\xi')}{(\phi_{2,i}^*,\phi _{2,i})_g}f(\xi ')g(\xi')d\xi'=
\int_0^1 \delta(\xi, \xi ') f(\xi')g(\xi ') d\xi'=f(\xi).  
\eew
\end{proof}
The generalized Fourier transform can be used to give the following interpretation 
of the Cauchy biorthogonal polynomials associated to discrete, finite positive measures 
$d\alpha, d\beta$.  
\bt {\bf Biorthogonal polynomials}
Let us consider the sequence 
$\chi _j:=\chi_{(\xi_{n-j}-\epsilon, \xi_{n-j}+\epsilon)}$ of indicator functions enclosing points $\xi_{n-j}$ of the support of the measure $g$ appearing in the cubic string equation \eqref{eq:cubicstring}, with $\epsilon$  small enough to ensure 
non overlapping supports.   Furthermore, let us consider the generalized Fourier transforms 
for  Types $a,b$ as in Theorem \ref{thm:isometry} and define: 
$$
p_j(x):=\hat \chi _{a,j}(x), \qquad q_j(y):=\hat \chi _{b,j}(y).  
$$ 
Then, $\{p_j, q_k\}$ are (un-normalized) Cauchy biorthogonal polynomials, that is 
$<\hat \chi_{a,i}|\hat \chi_{b,j}>=0,\, i\ne j$ and $<\hat \chi_{a,i}|\hat \chi_{b,i}>\ne 0$. 
\et 
\begin{proof} 
Both
$\hat \chi_{a,i}(x)$ and $\hat \chi_{b,j}(y)$ are polynomials in $x$, $y$
 respectively,  whose 
degrees are 
$\deg \hat
\chi_{a,j}(x)=\deg \hat
\chi_{b,j}(y)=j$  by \eqref{eq:Icubicstring*}.  The biorthogonality follows from 
Theorem \ref{thm:isometry}.  
\end{proof}

\section{ Acknowledgments} 
We would like to to thank R. Beals for sharing with us his insight regarding the Parseval 
Identity for an ordinary inhomogeneous string of M.G. Krein which influenced 
the formulation of Theorem \ref{thm:isometry}. 
\bibliographystyle{unsrt}
\bibliography{BOP}
\end{document}